# Giant enhancement of spin pumping efficiency using $Fe_3Si$ ferromagnet

Y. Ando[1], K. Ichiba[1], S. Yamada[2], E. Shikoh[1], T. Shinjo[1], K. Hamaya[2], M. Shiraishi[1]

[1]*Graduate School of Engineering Science, Osaka University, Osaka, Japan*

[2]*Department of Electronics, Kyushu University, Fukuoka, Japan*

*Corresponding author: E-mail: ando@ee.es.osaka-u.ac.jp

*Spincurrentronics*, **which involves the generation, propagation and control of spin currents, has attracted a great deal of attention because of the possibility of realizing dissipation-free information propagation. Whereas electrical generation of spin currents originally made the field of spincurrentronics possible, and significant advances in spin-current devices has been made, novel spin-current-generation approaches such as dynamical methods have also been vigorously investigated. However, the low spin-current generation efficiency associated with dynamical methods has impeded further progress towards practical spin devices. Here we show that by introducing a Heusler-type ferromagnetic material, $Fe_3Si$, pure spin currents can be generated about twenty times more efficiently using a dynamical method. This achievement paves the way to the development of novel spin-based devices.**

A central issue with respect to the practical application of spintronic devices is to establish a mechanism for highly efficient spin-current generation.[1] The development of magnetic tunnel junctions with single-crystal MgO barriers has addressed this challenge,[2,3] resulting in applications in magnetic heads and magnetoresistive random access memory. The electrical spin-injection from ferromagnetic materials (FM) into nonmagnetic materials (NM) is also in a similar situation.[4,5] The use of a tunnel barrier in order to overcome the conductance mismatch problem,[6] and the use of half-metal materials as spin injectors have also been proposed.[7] While steady progress has been made in the application of electrical methods for spin-current generation in spintronics devices, recent studies have also focused on more radical approaches such as dynamical,[8-16] thermal,[17] and acoustic[18] methods. These methods are expected to pave the way for a new generation of novel spintronics devices that involve no charge current. Spin pumping is a dynamical method in which a spin current is generated by a precession of the magnetization. It has been the subject of considerable interest because a spin current can be produced over a large area without the presence of a charge current, which is expected to reduce the problem of conductance mismatch.[13] Whereas, spin pumping is a promising technique for a next generation spin current devices, low efficiency of generation of pure spin current impedes further progress towards practical spin devices, unfortunately. For this reason, identifying a novel FM material that is capable of highly efficient spin injection is of the utmost importance. Here, we focus on single-crystal $Fe_3Si$, which has desirable properties such as a smaller damping constant and a larger resistivity than those for $Ni_{80}Fe_{20}$ (Py), the most commonly used spin source.[19] Moreover, high-quality single-crystal $Fe_3Si$ can be easily grown on semiconducting substrates such as Si, Ge and GaAs with atomically flat interfaces.[19-21] This means that $Fe_3Si$ can be applied to a wide variety of materials, allowing the development of novel semiconductor-based spintronic devices in addition to metal-based devices. In the present study, a significant enhancement of spin-injection efficiency is demonstrated by using a single-crystal $Fe_3Si$ layer.

A 25-nm-thick $Fe_3Si$ epitaxial layer was grown on a high-resistivity FZ-Si(111) substrate by molecular beam epitaxy (MBE) at room temperature.[19,20] A 5-nm-thick Pd layer was then formed by electron beam (EB) evaporation at room temperature. Two contact wires (separated by $w=1.0$ mm gap) for measuring the DC electromotive force were attached to the edge of the Pd film using Ag paste. During the measurements, microwaves with a frequency of 9.61±0.01 GHz were generated in a $TE_{102}$ cavity of an electron spin resonance (ESR) system, and an external static magnetic field, $H$, was applied at an angle, $\theta_H$, as shown in Fig. 1(a). The sample was placed inside the cavity in a nodal position where the rf electric and magnetic field components were a minimum and a maximum, respectively. The DC electromotive force, $V_{EMF}$, was measured using a nanovoltmeter. See the Methods section for details concerning device fabrication and the measurement setup. All measurements were carried out at room temperature.

Figure 1(b) shows ferromagnetic resonance (FMR) spectra, i.e., $dI(H)/dH$ as a function of $H-H_{FMR}$, for the Pd/Fe$_3$Si/Si sample recorded at $\theta_H = 0, 80, 110$, and 180°, where $I$, $H$, and $H_{FMR}$ are the microwave absorption intensity, external magnetic field, and FMR field, respectively. Unfortunately, FMR could not be measured at $\theta_H = 90°$ due to the limited external magnetic field strength, i.e., the maximum magnetic field of 1.3 T in the ESR system is smaller than the anisotropy field for the Fe$_3$Si thin film (~1.5 T). For $\theta_H = 0, 80, 110$, and 180°, clear FMR spectra were observed. From the obtained resonant magnetic field ($H_{FMR} = 92.9$ mT) at $\theta_H = 0°$, the saturation magnetization, $M_s$, is estimated to be 828 emu/cc, which is consistent with previously measured values using a vibrating sample magnetometer,[19,22] indicating that the spectra are associated with FMR in the Fe$_3$Si layer. Figure 1(c) shows $V_{EMF}$ as a function of $H-H_{FMR}$. For $\theta_H = 0°$, a clear signal can be seen at the FMR condition. The EMF signals were analyzed using a deconvoluted fitting function with independent contributions from the the inverse spin Hall effect (ISHE, symmetrical Lorentzian curve centered on $H_{FMR}$) and the anomalous Hall effect (AHE, asymmetrical curve) as follows:[11]

$$V_{EMF} = V_{ISHE} \frac{\Gamma^2}{(H-H_{FMR})^2+\Gamma^2} + V_{AHE} \frac{-2\Gamma(H-H_{FMR})}{(H-H_{FMR})^2+\Gamma^2} , \qquad (1)$$

where $\Gamma$ is the damping constant. As shown in Fig. 1(d), a theoretical fit using Eq. (1) nicely reproduces the experimental results. $V_{ISHE}$ and $V_{AHE}$ are estimated to be 67.1 and 17.5 µV/mm, respectively. Figure 1(e) shows $V_{ISHE}$ and $V_{AHE}$ as a function of $\theta_H$. The polarity reversal observed for $V_{ISHE}$ when $\theta_H$ is changed from 0° to 180° is consistent with the theoretically predicted symmetry of the ISHE, expressed as $J_c = J_s \times \sigma$, where $\sigma$, $J_s$ and $J_c$ are the directions of the spin, spin current and charge current,[11] respectively, thus indicating successful dynamical spin injection into the Pd layer from the $Fe_3Si$ layer. This is also supported by the linear relationship between $V_{ISHE}$ and the microwave power, $P_{MW}$, shown in the inset of Fig. 1(f) (see also Supplementary Information (SI) **A**). Since the conductances of the Pd and $Fe_3Si$ layers are in parallel to each other, the electromotive force generated in the $Fe_3Si$ layer is also detected. Although the anisotropic magnetoresistance (AMR) effect can produce signals with a Lorentzian line shape in the $V_{EMF}$-$H$ curve,[23] the $\theta_H$ dependence of $V_{ISHE}$ induced by the AMR is quite different from that shown in Fig. 1(e). In addition, no such Lorentzian line shape was obtained for the $Fe_3Si$ layer in the absence of the Pd layer. Considering these results, it can be concluded that the contribution of the AMR effect is negligibly small (see SI **B**). Furthermore, although in the FMR condition, a temperature gradient is induced in the sample, and this can lead to an additional DC electromotive force due to the Seebeck effect, the spin Seebeck effect,[17] and the anomalous Nernst-Ettingshausen effect,[24,25] these contributions were also found to be negligible (see SI **B**). Therefore, it can be concluded that the origin of $V_{ISHE}$ is the ISHE in the Pd layer due to a pure spin current generated by spin pumping of the $Fe_3Si$ layer. In fact, when the NM layer was changed from Pd to Al, in which spin-orbit interactions are weaker than in Pd, $V_{ISHE}$ was drastically reduced to 4.20 µV/mm, which is one-sixteenth of the value for the Pd/$Fe_3Si$/Si sample (see SI **C**).

For comparison, the spin injection efficiency was investigated for several FM materials: $Ni_{80}Fe_{20}$ (Py), polycrystalline $Fe_3Si$, and single-crystal $Co_6Fe_4$. The polycrystalline Py and $Fe_3Si$ layers were formed by EB evaporation and pulse laser deposition, respectively. The single-crystal $Co_6Fe_4$ was grown by MBE.[26] The detailed growth procedures are described in the Methods section. To distinguish between the single-crystal $Fe_3Si$ grown by MBE and the polycrystalline $Fe_3Si$ grown by PLD, these layers are referred to as "single-$Fe_3Si$" and "poly-$Fe_3Si$", respectively. Figure 2 shows the $H$ dependence of the (top) FMR signal, $dI(H)/dH$, and the (bottom) electromotive force, $V_{EMF}$, for $\theta_H = 0$ and $180°$, for a) Pd/Py/$SiO_2$/Si, b) Pd/poly-$Fe_3Si$/$SiO_2$/Si, and c) Pd/$Co_6Fe_4$/Si. The microwave excitation power was 200 mW. Clear FMR signals and EMFs were obtained for all samples. In order to estimate the generated spin current, $J_s^0$, the following equation was used:[12]

$$\frac{V_{ISHE}}{w} = \frac{\theta_{SHE}\lambda_N tanh\left(\frac{d_N}{2\lambda_N}\right)}{d_N\sigma_N + d_F\sigma_F}\left(\frac{2e}{\hbar}\right)J_s^0 , \qquad (2)$$

where $d_F$ and $\sigma_F$ are the thickness and electric conductivity of the FM layer, and $d_N$ and $\sigma_N$ are those of the Pd layer, respectively. From the $V_{EMF}$ vs. $H$ curves, $V_{ISHE}/w$ for the Pd/Py, Pd/poly-$Fe_3Si$, and Pd/$Co_6Fe_4$ samples was estimated to be 2.85, 15.0, and 2.92 µV/mm, respectively. This leads to the surprising conclusion that $V_{ISHE}/w$ for the single-$Fe_3Si$ sample (67.1 µV/mm) is more than twenty times higher than that for samples using a conventional FM material such as Py. From Eq. (2), $J_s^0$ for the single-$Fe_3Si$, poly-$Fe_3Si$, Py, and $Co_6Fe_4$ samples is calculated to be $2.75\times10^{-8}$, $5.76\times10^{-9}$, $1.25\times10^{-9}$, and $1.76\times10^{-9}$ J/m$^2$, respectively (see Table 1). Thus, for the single-$Fe_3Si$ sample, the generated spin current is more than twenty times higher than that for the Py samples. Since $J_s^0$ is a good indicator of the spin injection efficiency, these results clearly indicate that highly efficient spin injection is realized for the single-$Fe_3Si$ sample. In general, $J_s^0$ is expressed as[12]

$$J_s^0 = \frac{g_r^{\uparrow\downarrow}r^2h^2\hbar\left[4\pi M_s\gamma + \sqrt{(4\pi M_s)^2\gamma^2 + 4\omega^2}\right]}{8\pi\alpha^2[(4\pi M_s)^2 + 4\omega^2]} , \qquad (3)$$

where $h$, $\hbar$, $g_r^{\uparrow\downarrow}$, $M_s$ and $\alpha$ are the microwave magnetic field, the Dirac constant, the real part of the mixing conductance, the saturation magnetization and the Gilbert damping constant, respectively. $\omega$ ($=2\pi f$) is the angular frequency of the magnetization precession, where $f$ is the microwave frequency.

The estimated $g_r^{\uparrow\downarrow}$ values and other physical parameters for the different samples are summarized in Table 1. The parameters α and $M_s$ are estimated from width of FMR spectrum and $H_{FMR}$, respectively. These parameters are strongly dependent on the FM layer, and Eq. (3) implies that for high spin injection efficiency, α should be as small as possible and $M_s$ should be optimized to maximize $J_s^0$. However, even though the α values for the poly-Fe$_3$Si sample is smaller than that for the single-Fe$_3$Si sample, and the $M_s$ value is comparable, $J_s^0$ for the poly-Fe$_3$Si sample is considerably smaller than that for the single-Fe$_3$Si sample. This indicates that α and $M_s$ are not the main factors responsible for the large $V_{ISHE}$ for the single-Fe$_3$Si sample. We therefore focus on $g_r^{\uparrow\downarrow}$, which is generally related to the conductance between non-collinear FMs. In this study, since $g_r^{\uparrow\downarrow}$ is calculated using Eq. (3), other extrinsic contributions that affect the spin injection efficiency, which are not considered in the conventional theory, are also included in $g_r^{\uparrow\downarrow}$. As can be seen from Table 1, $g_r^{\uparrow\downarrow}$ for the single-Fe$_3$Si sample is clearly larger than those for the other samples. It should also be noted that $g_r^{\uparrow\downarrow}$ for the Co$_6$Fe$_4$ sample is also relatively large despite the small $J_s^0$ value. Both of these results were reproducible over several samples. This unexpected behavior of $g_r^{\uparrow\downarrow}$ might provide an important clue for understanding the mechanism that gives rise to the large $V_{ISHE}$ for the single-Fe$_3$Si sample.

Based on the results shown in Fig. 2 and Table 1, a possible mechanism is now considered. Figure 3 schematically illustrates the spin-current flow generated by spin pumping in samples consisting of NM and FM layers, for different interface conditions between the FM layer and the substrate. Figure 3(a) shows the case for a single FM layer with an atomically flat interface with the substrate, Fig. 3(b) shows the case for a rough interface between a single FM layer and the substrate, and Fig. 3(c) shows a situation where there are two different FM layers with different saturation magnetizations, and an atomically flat interface exists between the lower FM2 layer and the substrate. Although the ideal spin pumping condition is represented by Fig. 3(a), it is possible that a non-zero interface rougness exists between FM1 layer and substrate. In this case, since ferromagnetic resonance condition of the magnetization near the interface is changed due to shape anisotropy,[27] magnetization near the interface does not precess under the same FMR condition as that for the FM1 layer. Since the spin diffusion length in FM1 layer is short, the FM1 layer near the interface in Fig. 3(b) can act as an effective spin sink, resulting in a reduction of the spin current flowing into the NM layer. The FM2 layer also induces an additional EMF, which can cause a significant change in the measured $V_{\text{ISHE}}$. A similar situation should occur for the sample shown in Fig. 3(c), with atomically flat FM2 layer. In the present study, Fig. 3(a) corresponds to the case for the single-$Fe_3Si$ and the $Co_6Fe_4$ samples, and Fig. 3(b) corresponds to the case for the Py and poly-$Fe_3Si$ samples because the thermally oxidized Si substrate has a non-zero surface roughness. In fact, a measureable $V_{\text{ISHE}}$ was found for a Py layer without any NM layer, which indicates the existence of another spin sink.[28] Although the $Co_6Fe_4$ sample also has an ideal atomically flat interface, across which spins can be electrically injected from the $Co_6Fe_4$ into a Si channel even at room temperature,[29, 30] the large $\alpha$ and $M_s$ values lead to a significant reduction in the spin-current density, as indicated by Eq. (3). Thus, $g_r^{\uparrow\downarrow}$ might be also reflect the crystal and magnetic quality of the ferromagnetic layer.

To experimentally investigate whether this was in fact the case, two additional single-$Fe_3Si$ samples were fabricated with different interfacial conditions. First, an as-grown single-$Fe_3Si$/Si sample was annealed at 350 °C for 30 min in an Ar atmosphere because an earlier study revealed that slight intermixing occurs between the $Fe_3Si$ layer and the Si substrate at around 300 °C.[22] Thus, the annealed sample expected to have a rough interfacial layer, which corresponds both to the sample in Fig. 3(b) and that in Fig. 3(c). Following annealing, a 5-nm-thick Pd layer was formed. FMR spectra for the as-grown and annealed samples are shown in Fig. 4(a). To highlight the differences in the FMR field between these two samples, the microwave absorption intensity, $I$, rather than $dI/dH$, is plotted as a function of $H$-$H_{\text{FMR}}$. As can be seen, the width of the FMR spectrum for the annealed sample is slightly larger. As indicated by the blue arrow, the absorption intensity under a high external magnetic field is enhanced for the annealed sample, which indicates the presence of a Si-rich interfacial layer. Figure 4(b) shows $V_{EMF}$ against $H\text{-}H_{FMR}$ for the annealed sample, measured at $\theta_H = 0°$ and $180°$. The microwave excitation power was 200 mW. The signal shape is seen to be significantly different to that for the as-grown sample shown in Fig. 1(c). The magnitude of $V_{ISHE}$ was estimated to be 12.6 µV/mm, which is about one-fifth of that for the as-grown sample. To investigate the effect of an intentionally inserted additional FM layer near the interface with the substrate, as in Fig. 3(c), two further samples were fabricated. In these samples, a 2-nm-thick layer of either $Fe_4Si$ or $Fe_2Si$ was grown on the substrate before growth of the single-$Fe_3Si$ layer. Despite the large composition change, no evidence was found that the presence of such an interfacial layer affected the epitaxial growth of the single-$Fe_3Si$ layer, and the interfaces remained atomically flat. Figure 4(c) shows $V_{EMF}$ against $H\text{-}H_{FMR}$ for these two samples. The magnitude of $V_{ISHE}$ was drastically reduced to 13.2 µV/mm for the sample with $Fe_4Si$ and 16.2 µV/mm for the sample with $Fe_2Si$, despite the presence of atomically flat interfaces. The results shown in Fig. 4 strongly support the idea that to realize highly efficient spin injection using spin pumping techniques, magnetic properties of the FM layer near the interface should be carefully considered. These findings are likely to have a major impact in the field of spintronics. We believe that further enhancement of the spin injection efficiency can be realized by using completely uniform FM metals with a much smaller $\alpha$ value, such as Co-based Heusler alloys.[26]

**Methods**

Undoped Si(111) wafers were used as substrates for growing single-crystal $Fe_3Si$, and $Co_6Fe_4$ layers. After cleaning the substrate with an aqueous HF solution (HF:$H_2O$=1:40), a heat treatment was carried out at 450 °C for 20 min in a molecular beam epitaxy (MBE) reaction chamber with a base pressure of $2\times10^{-9}$ Torr. Transmission electron microscopy observations revealed that interfaces in the $Fe_3Si$ and $Co_6Fe_4$ samples were atomically flat. Polycrystalline Py and $Fe_3Si$ layers were formed on thermally oxidized Si(100) substrates (oxide thickness 500 nm) using electron beam evaporation and pulse laser deposition, respectively. After the substrates were cleaned with acetone and isopropanol, a 25-nm-thick Py or $Fe_3Si$ layer was formed at room temperature. After deposition of the ferromagnetic layer, a polycrystalline nonmagnetic layer such as Pd and Al was formed using electron beam evaporation at room temperature. The dimensions of the samples were 2 mm × 1 mm.

Two lead wires for measuring the electromotive force were attached to the edge of the nonmagnetic layer using Ag paste. The sample was placed near the center of a $TE_{102}$ microwave cavity in an electron spin resonance (ESR) system (Bruker EMX10/12), where the magnetic-field component was a maximum and the electric-field component was a minimum. Microwaves with a frequency of 9.61±0.01 GHz, and a static external magnetic field were applied to the samples. The electromotive force was measured using a nanovoltmeter (KEITHLEY 2182A) and all measurements were performed at room temperature.

**Acknowledgements**

This research was supported in part by a Grant-in-Aid for Scientific Research from the MEXT, Japan, by STARC, by the Adaptable & Seamless Technology Transfer Program through Target-driven R&D from JST, and by the Toray Science Foundation.

**Additional information**

The authors declare no competing financial interests.

**Figure legends**

**Figure 1 | Electromotive force measurements for Pd/Fe$_3$Si/Si sample.**

**a)** Schematic illustration of the Pd/Fe$_3$Si/Si sample structure. The lateral dimensions of the Fe$_3$Si layer were 2 mm (*w*) × 1 mm and the thickness, *d*, was 25 nm. Two contact wires were attached to the Pd layer using Ag paste. The electrode separation, *w*, was 1.0 mm. The static external magnetic field, *H*, was applied at an angle of $\theta_H$ to the Fe$_3$Si film plane. **b)** FMR spectra, d*I*(*H*)/d*H*, for the Fe$_3$Si sample at $\theta_H$ = 0, 80, 110 and 180° as a function of *H*-*H*$_{FMR}$, where *I* is the microwave absorption intensity in arbitrary units. The microwave power was 200 mW. The FMR field, *H*$_{FMR}$, for $\theta_H$=0° was estimated to be 92.9 mT. **c)** Dependence of the electromotive force, *V*$_{EMF}$, on *H* for $\theta_H$=0, 80, 110 and 180°. **d)** Dependence of the electromotive force, *V*$_{EMF}$, on *H* for $\theta_H$=0°. The open circles are experimental data, and the green solid line is a fit obtained using Eq. (1) considering the contributions from the ISHE and AHE. The red and blue lines are fits for the ISHE signal from the Pd layer and the AHE signal from the Fe$_3$Si layer, respectively. **e)** Dependence of *V*$_{ISHE}$ and *V*$_{AHE}$ on the magnetic field angle, $\theta_H$, where *V*$_{ISHE}$ and *V*$_{AHE}$ are the electromotive forces due to the ISHE and the AHE, respectively. **f)** Dependence of *V* on *H* for different microwave powers at $\theta_H$=0°. The inset shows the microwave power dependence of *V*$_{ISHE}$ and *V*$_{AHE}$.

**Figure 2 | Electromotive force measurements for different ferromagnetic samples.**

The Ni$_{80}$Fe$_{20}$ (Py) and polycrystalline Fe$_3$Si layers were deposited on thermally oxidized Si(100) substrates (oxide thickness 500 nm) using electron beam evaporation and pulse laser deposition, respectively, at room temperature. The single-crystal Co$_6$Fe$_4$ layer was grown by molecular beam epitaxy at room temperature. The microwave power was 200 mW.

$H$ dependence of the (top) FMR signal, $dI(H)/dH$, and the (bottom) electromotive force, $V_{EMF}$, at $\theta_H=0°$ and $180°$ for **a)** Pd/Py/SiO$_2$/Si, **b)** Pd/poly-Fe$_3$Si/SiO$_2$/Si, and **c)** Pd/Co$_6$Fe$_4$/Si samples. The microwave excitation power was 200 mW. The FMR and EMF measurement procedures and the sample geometry were the same as those for the Pd/single-crystal Fe$_3$Si/Si sample shown in Fig. 1(a). $H_{FMR}$ was estimated to be 131.6, 89.9, and 51.2 mT for the Py, poly-Fe$_3$Si, and Co$_6$Fe$_4$ sample, respectively.

**Table 1 | Summary of physical parameters.**

Physical parameters for estimating $J_s^0$ and $g_r^{\uparrow\downarrow}$ for different ferromagnetic samples. $M_s$ and $\alpha$ were obtained from $H_{FMR}$ and the linewidth of the FMR spectrum, respectively. The conductivity and spin diffusion length for the Pd layer are $4.08\times10^6$ $\Omega^{-1}m^{-1}$ and 9 nm, respectively, as reported in Ref. 28.

**Figure 3| Schematic illustration of spin-current flow under FMR conditions for samples with different interface structures.**

Different interface structures between FM layer and substrate, **a)** an atomically flat interface, **b)** rough interface, and **c)** an atomically flat interface with an interfacial FM2 layer whose saturation magnetization is different from that of the FM1 layer. The schematics show the spin current flow under FMR conditions for the FM1 layer. The upper figure represents ideal spin pumping conditions.

**Figure 4| Electromotive force measurements for Pd/Fe$_3$Si/Si samples with different interfacial layers.**

**a)** FMR spectra for as-grown Pd/Fe$_3$Si/Si sample and sample annealed at 350 °C for 30 min in an Ar atmosphere. Slight intermixing between the $Fe_3Si$ layer and Si substrate is known to occur at around 300 °C. After annealing, a 5-nm-thick Pd layer was deposited. In order to highlight the differences in the FMR field between these two samples, the microwave absorption intensity, $I$, rather than $dI/dH$, is plotted as a function of $H-H_{FMR}$. $H_{FMR}$ for the as-grown and annealed samples is estimated to be 88.0 and 87.7 mT, respectively. **b)** DC EMF for the annealed sample at $\theta_H = 0°$ and 180°. The microwave excitation power was 200 mW. The EMF measurement procedure and the sample geometry were the same as those for the Pd/single-$Fe_3Si$/Si sample. The open circles are experimental data and the solid line is a fit obtained using Eq. (1). $V_{ISHE}$ and $V_{AHE}$ were estimated to be 12.6 and 10.4 µV/mm, respectively. **c)** DC EMF for Pd/$Fe_3Si$(23 nm)/$Fe_{3-X}Si_{1+X}$(2 nm)/Si samples at $\theta_H = 0°$. The microwave excitation power was 200 mW. The composition of the 2-nm-thick interfacial layer is either $Fe_4Si$ (blue) or $Fe_4Si$ (green). The interfaces were confirmed to be atomically flat. A representative $V_{EMF}$-$H$ curve for the Pd/$Fe_3Si$(25 nm)/Si sample is displayed using red circles. The open circles are experimental data and the solid line is a fit obtained using Eq. (1).

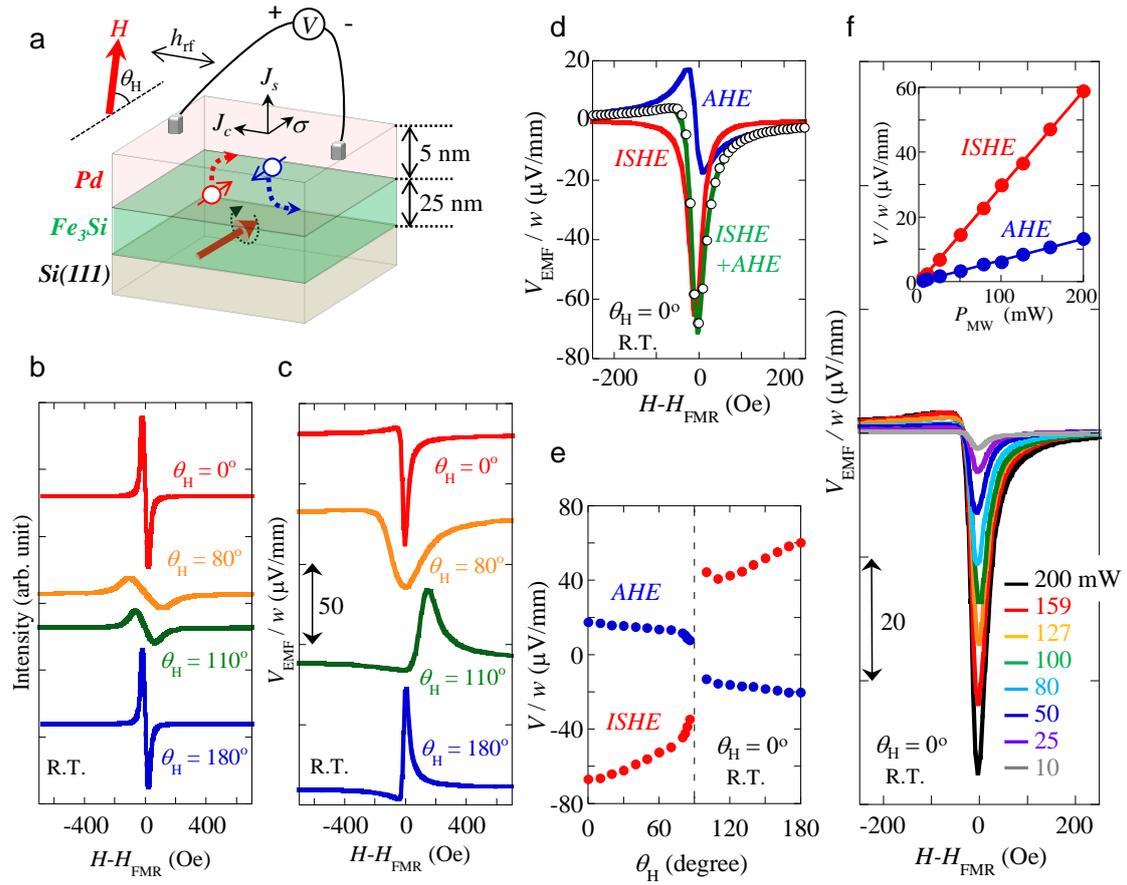

Fig.1 Y. Ando

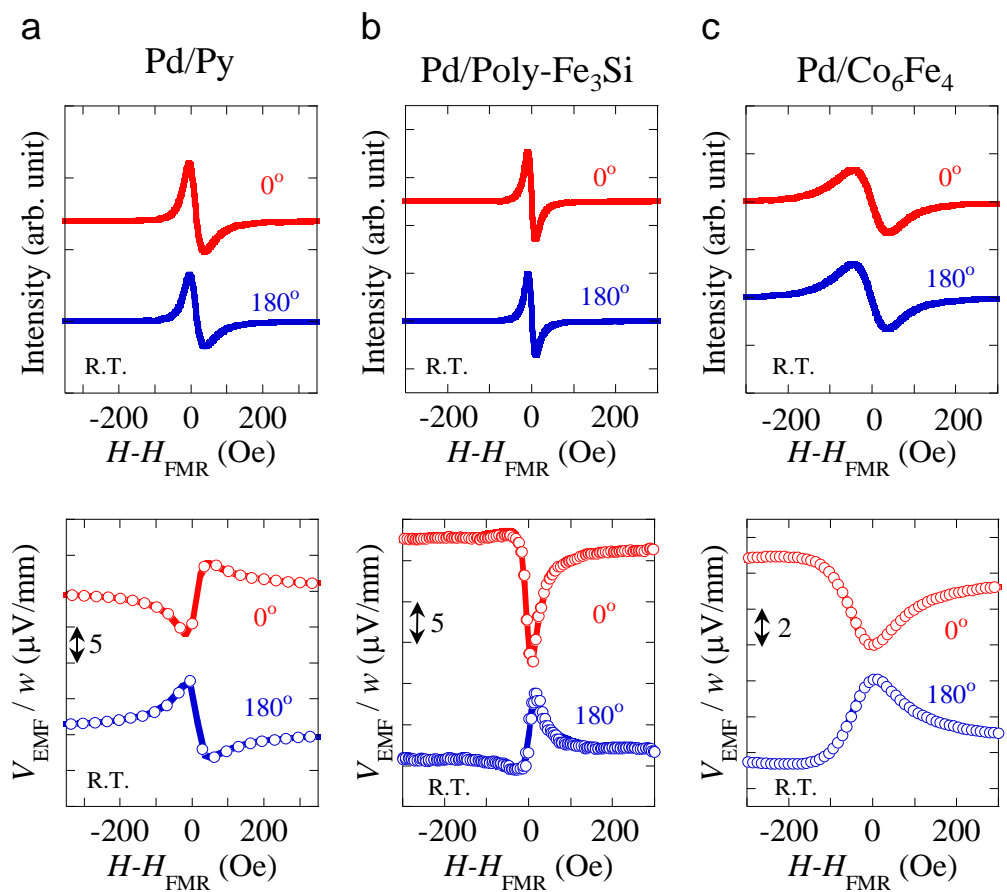

Fig.2 Y. Ando

# Table 1 Y. Ando

| | | $M_s$ (emu/cc) | $\alpha$ | $\sigma_F$ ($\Omega^{-1}$m$^{-1}$) | $J_s^0$ (J/m$^2$) | $g_r^{\uparrow\downarrow}$ (m$^{-2}$) |
|---|---|---|---|---|---|---|
| Fe$_3$Si | Single crystalline | 828 | 0.0087 | $1.3 \times 10^6$ | $2.75 \times 10^{-8}$ | $6.2 \times 10^{20}$ |
| Fe$_3$Si | Polycrystalline | 860 | 0.0050 | $1.3 \times 10^6$ | $5.76 \times 10^{-9}$ | $2.5 \times 10^{19}$ |
| Py | Polycrystalline | 535 | 0.0149 | $2.5 \times 10^6$ | $1.25 \times 10^{-9}$ | $5.2 \times 10^{19}$ |
| Co$_6$Fe$_4$ | Single crystalline | 1600 | 0.0227 | $5.0 \times 10^6$ | $1.76 \times 10^{-9}$ | $3.1 \times 10^{20}$ |

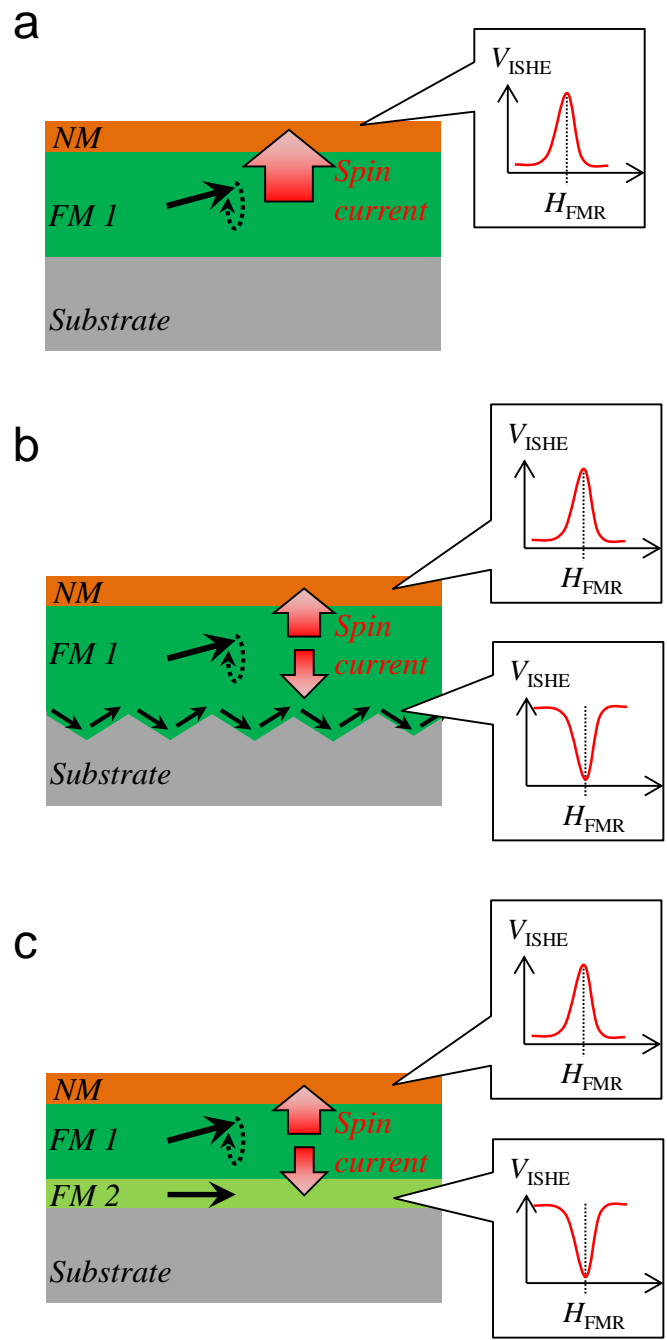

Fig.3 Y. Ando

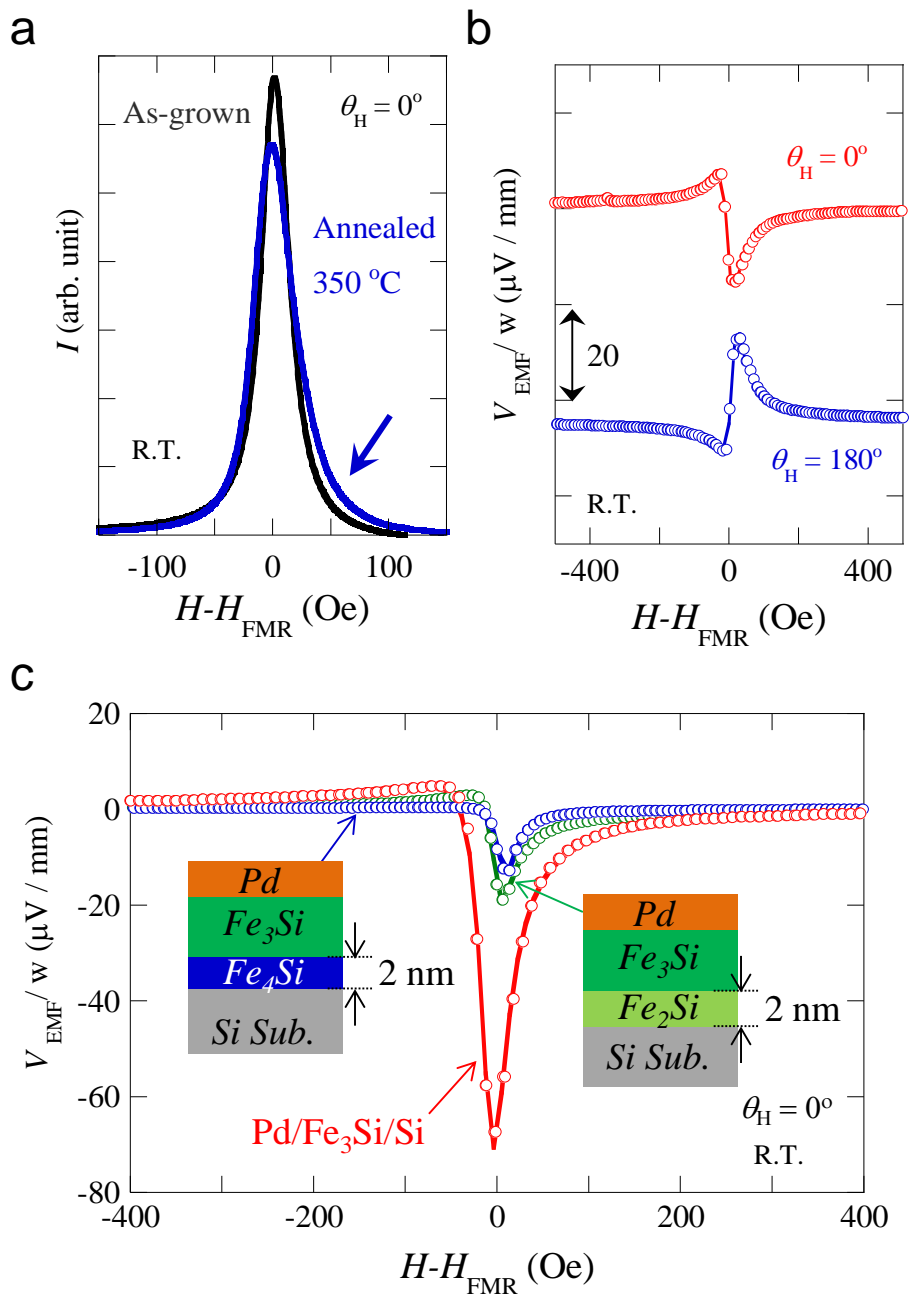

Fig.4 Y. Ando

*Supplementary Information*

# Giant enhancement of spin pumping efficiency from Fe$_3$Si ferromagnet

Y. Ando[1,#], K. Ichiba[1], S. Yamada[2], E. Shikoh[1], T. Shinjo[1], K. Hamaya[2], M. Shiraishi[1]

[1] *Graduate School of Engineering Science, Osaka University, Osaka, Japan*

[2] *Department of Electronics, Kyushu University, Fukuoka, Japan*

**A. Theoretical prediction of the microwave power dependence of $V_{ISHE}$**

As expressed in Eqs. (1)–(3), $V_{ISHE}$ has an $h^2$ dependence, where $h$ is the microwave magnetic field. Since $h$ has a linear relationship with $\sqrt{P_{MW}}$, $V_{ISHE}$ is expected to increase linearly with $P_{MW}$. Therefore, the existence of a linear relationship between the microwave power and $V_{ISHE}$ is a reliable indicator of successful spin pumping.

**B. Contribution of spurious effects**

Since it is possible that extrinsic EMFs unrelated to the inverse spin Hall effect in the Pd layer may also be detected using our experimental setup, these need to be taken into consideration. For this reason, control experiments were carried out to investigate the anisotropic magnetoresistance (AMR) effect, the Seebeck effect, the spin Seebeck effect, and the anomalous Nernst-Ettingshausen effect.[1-3] The AMR effect produces signals with Lorentzian line shapes in the $V_{EMF}$-$H$ curve when a charge current is generated in the FM layer. The main origin of the charge current is induction due to the microwave and external magnetic fields.[4-6] In order to investigate the contribution of the AMR effect, the EMF was first measured for a single-Fe$_3$Si layer without a Pd layer, as shown in the inset of Fig. S1(a). $V_{EMF}$–$H$ curves for $\theta_H = 0°$ and 180° are shown in the main panel of Fig. S1(a). The measurements were performed at room temperature and the microwave excitation power was 200 mW. As can be seen, the signal shapes are quite different from those for the Pd/single-Fe$_3$Si/Si sample shown in Fig. 1(c). $V_{ISHE}$ and the $V_{ISHE}/V_{AHE}$ ratio are estimated to be 2.7 µV/mm and 0.35 respectively, indicating that the intensity of signals with a Lorentzian shape is drastically reduced due to absence of the Pd layer. If $V_{ISHE}$ for the Pd/Fe$_3$Si sample was mainly due to the AMR effect in the Fe$_3$Si layer, $V_{ISHE}$ should remain the same or increase in the absence of a Pd layer. It can therefore be concluded that the contribution of an EMF due to the AMR effect is negligibly small. The clear difference in the $\theta_H$ dependence of $V_{ISHE}$ also supports this conclusion.

The contributions of the spin Seebeck and anomalous Nernst-Ettingshausen effects are next considered. These effects occur when a vertical thermal gradient exists under the FMR condition, as shown in Fig. S1(d).[1-3] In fact, the $\theta_H$ dependence of the EMF induced by the anomalous Nernst-Ettingshausen effect or the spin Seebeck is the same as that for the inverse spin Hall effect. Here, the EMF is compared for the Pd/Fe$_3$Si/Si, Fe$_3$Si/Si and Al/Fe$_3$Si/Si samples. The thermal conductivities of Pd, Al, air and Si are reported to be 71.8, 200, 0.026, and 149 Wm$^{-1}$K$^{-1}$,[7-10] respectively. Therefore, the vertical thermal gradient in the Pb/Fe$_3$Si/Si sample is expected to be smaller than that in the Fe$_3$Si/Si sample and to be in the opposite direction to that in the Al/Fe$_3$Si/Si sample. If the anomalous Nernst-Ettingshausen effect or the spin Seebeck effect had a dominant influence on $V_{ISHE}$, $V_{EMF}$ for the Pd/Fe$_3$Si/Si sample should be smaller than that for the Fe$_3$Si/Si sample and should have the opposite polarity to that for the Al/Fe$_3$Si/Si sample. However, as shown in Fig. S1(b), $V_{ISHE}$ for the Al/Fe$_3$Si/Si sample was significantly smaller than that for the Pd/Fe$_3$Si/Si sample, and almost the same as that for the Fe$_3$Si/Si sample. This cannot be explained if $V_{ISHE}$ is mainly due to the anomalous Nernst-Ettingshausen effect or the spin Seebeck effect. Only the inverse spin Hall effect can produce such behavior because spin-orbit interactions are much weaker in Al than in Pd.

Finally, the contribution of the Seebeck effect is considered. This plays an important role when a lateral thermal gradient exists, as shown in Fig. S1(e). A sample was fabricated in which a Pd layer was deposited only in the contact area, as shown in the inset Fig. S1(c). Since the Pd layer between the contacts is missing, no ISHE occurs but any contribution from the Seebeck effect would not be affected. As shown in Fig. S1(c), $V_{ISHE}$ is estimated to be 2.4 µV/mm, which is small enough to conclude that the Lorentzian signal observed in the Pd/Fe$_3$Si/Si sample is not due to the Seebeck effect.

### C. Electromotive force in the Al/Fe$_3$Si/Si sample

Since the conductivity of the Al layer is $9.42 \times 10^6\, \Omega^{-1} m^{-1}$, which is more than twice as large as that of the Pd layer ($4.08 \times 10^6\, \Omega^{-1} m^{-1}$), it is necessary to take into account the change in conductivity of the entire sample, $d_N \sigma_N + d_F \sigma_F$, in Eq. (2) that occurs when the NM layer is changed. The $d_N \sigma_N + d_F \sigma_F$ values for the Al/single-Fe$_3$Si/Si and Pd/single-Fe$_3$Si/Si samples are calculated to be 0.125 and 0.060 $\Omega^{-1}$, respectively. On the other hand, the $V_{ISHE}$ values are estimated to be 4.20 and 67.1 µV/mm, respectively. Such a large discrepancy cannot be explained only in terms of the high conductivity of the Al layer. Therefore, it is clear that $V_{ISHE}$ strongly depends on the strength of spin-orbit interactions in the NM layer.

**Figure legends**

**Figure S1 | Control experiments for evaluating contribution of spurious effects.**

$H$ dependence of the electromotive force when $\theta_H=0°$ and $180°$ for **a)** single-Fe$_3$Si/Si sample with no Pd layer, **b)** 10-nm-thick Al/single-Fe$_3$Si/Si sample, and **c)** single-Fe$_3$Si/Si sample with Pd contacts. The measurements were performed at room temperature. The microwave excitation power was 200 mW. The FMR and EMF measurement procedures and the sample geometry were the same as those for the Pd/Fe$_3$Si/Si sample. Schematic illustrations of a FMR-induced thermal gradient along the **d)** vertical direction and **e)** lateral direction.

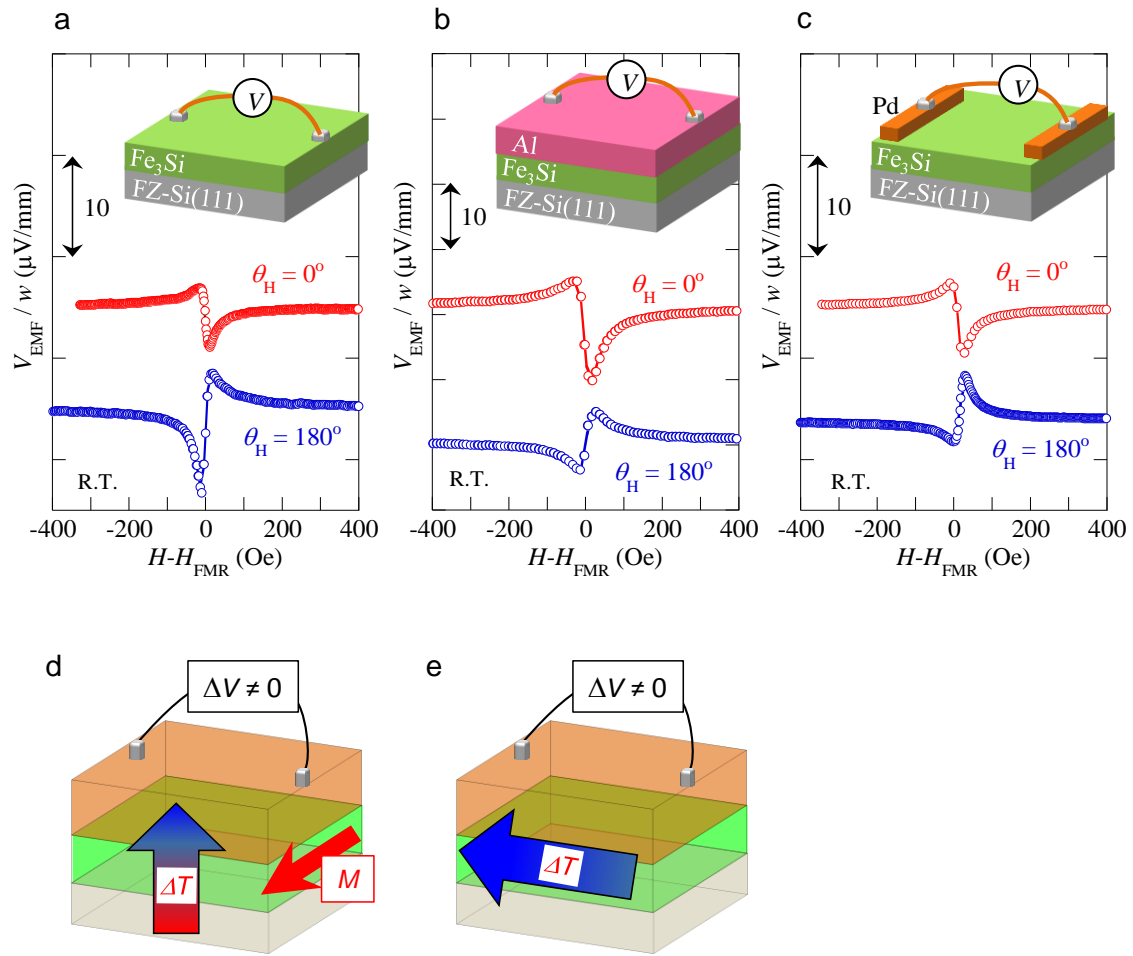

Fig.S1 Y. Ando